\documentclass{IEEEtran}
\usepackage{cite}
\usepackage{amsmath,amssymb,amsfonts}
\usepackage{graphicx}
\usepackage{subcaption}
\usepackage{textcomp,nicefrac}
\def\BibTeX{{\rm B\kern-.05em{\sc i\kern-.025em b}\kern-.08em
		T\kern-.1667em\lower.7ex\hbox{E}\kern-.125emX}}
\begin{document}
	\title{Feasibility study of a novel thermal neutron detection system using event mode camera and LYSO scintillation crystal}
	\author{Tianqi Gao, Mohammad Alsulimane, Sergey Burdin, Gabriele D'Amen, Cinzia Da Via, Konstantinos Mavrokoridis, Andrei Nomerotski, Adam Roberts, Peter Svihra, Jon Taylor, and Alessandro Tricoli
		\thanks{This work was supported in part by the NuSec-NNSA Research Collaboration Grants of the Nuclear Security Science Network and The University of Surrey.}
		\thanks{T. Gao and C. Da Via are with the School of Physics and Astronomy, The University of Manchester, Manchester M13 9PL, the UK (e-mail: tg514@cam.ac.uk).}
		\thanks{C. Da Via is also with the Physics and Astronomy Department, Stony Brook University 100 Nicolls Rd, Stony Brook, NY 11794, USA (e-mail: cinzia.da.via@cern.ch).}
		\thanks{G. D'Amen, A. Nomerotski, A. Tricoli is with Brookhaven National Laboratory, Upton, NY 11973, USA.}
		\thanks{M. Alsulimane, S. Burdin, K. Mavrokoridis, A. Roberts, J. Taylor is with the Department of Physics, University of Liverpool, Liverpool L1 8JX, the UK.}
		\thanks{P. Svihra is with CERN, 1211 Geneva 23, Switzerland.}}

	\maketitle
	
	\begin{abstract}
		The feasibility study of a new technique for thermal neutron detection using a Timepix3 camera (TPX3Cam) with custom-made optical add-ons operated in event-mode data acquisition is presented. The camera has a spatial resolution of $\sim$ 16 \textmu m and a temporal resolution of 1.56 ns. Thermal neutrons react with $^6$Lithium to produce a pair of 2.73 MeV tritium and 2.05 MeV alpha particles, which in turn interact with a thin layer of LYSO crystal to produce localized scintillation photons. These photons are directed by a pair of lenses to an image intensifier, before being recorded by the TPX3Cam. The results were reconstructed through a custom clustering algorithm utilizing the Time-of-Arrival (ToA) and geometric centre of gravity of the hits. Filtering parameters were found through data analysis to reduce the background of gamma and other charged particles. The efficiency of the converter is 4\%, and the overall detection efficiency of the system including the lead shielding and polythene moderator is $\sim$ 0.34\%, all converted thermal neutrons can be seen by the TPX3Cam. The experiment used a weak thermal neutron source against a large background, the measured signal-to-noise ratio is 1/67.5. Under such high noise, thermal neutrons were successfully detected and predicted the reduced neutron rate, and matched the simulated rate of the thermal neutrons converted from the source. This result demonstrated the excellent sensitivity of the system.
	\end{abstract}
	
	
	\section{Introduction}
	\label{sec:introduction}
	\IEEEPARstart{T}{hermal} neutron detection has been well studied over the years, with the ongoing development of radiation detectors, naturally comes the investigation of real-time detection with enhanced sensitivity of spatial and temporal resolutions. The sectors that have an interest in a neutron camera are reactor instrumentation, material science, and neutron imaging, amongst other applications. Thermal neutrons are usually not detected directly but with their reaction products, which are low-energy charged particles. These are achieved either by a sensor doped with a converter such as $^{10}$B and gadolinium-doped micro-channel plates (MCPs) as shown in \cite{tremsin2011improved}, or by depositing a converter on the surface of a sensitive volume as demonstrated in~\cite{jakubek2009neutron}, $^{6}$Li or $^{10}$B for thermal neutrons and polyethylene for fast neutrons. Similarly, thermal neutrons were detected by utilizing a $^{6}$LiF-coated silicon sensor tracking converted alpha or tritium ions in~\cite{krejci2016development}. Both of~\cite{jakubek2009neutron,krejci2016development} used the Timepix hybrid detector. Most recently,~\cite{losko2021new} used a technique of scintillating photons from the ions produced by the neutron converter $^{6}$LiF:ZnS, then used a silicon sensor Timepix3 detector coupled with an image intensifier to detect the photons.
	
	Thermal neutron conversion methods such as $^{10}$B(n,$\alpha$) reaction, $^{6}$Li(n,$\alpha$) reaction, or $^{3}$He(n,p) reaction are formulated below,
	\begin{subequations}\label{eq:reactions}
		\begin{equation}
			\label{eq:reaction:1}
			^{10}_{5}B + ^{1}_{0}n  \longrightarrow ^{7}_{3}Li + ^{4}_{2}\alpha,
			\\
		\end{equation}
		With the product energies:
		\begin{equation}
			\label{eq:reaction:2}
			E_{Li} = 0.84 \textrm{ MeV and } E_{\alpha} = 1.47\textrm{ MeV}.
		\end{equation}
		\begin{equation}
			\label{eq:reaction:3}
			^{6}_{3}Li + n  \longrightarrow ^{3}_{1}H + \alpha
			\\
		\end{equation}
		With the product energy of:
		\begin{equation}
			\label{eq:reaction:4}
			E_{H} = 2.73 \textrm{ MeV and } E_{\alpha} = 2.05\textrm{ MeV}.
		\end{equation}
		\begin{equation}
			\label{eq:reaction:5}
			^{3}_{2}He + n  \longrightarrow ^{3}_{1}H + p.
			\\
		\end{equation}
		With the product energy of:
		\begin{equation}
			\label{eq:reaction:6}
			E_{H} = 0.573 \textrm{ MeV and } E_{p} = 0.191\textrm{ MeV}.
		\end{equation}
	\end{subequations}
	
	The energy of the reaction products in \eqref{eq:reaction:4} are the highest in the available options.
	
	A report on a fast $\alpha$-imaging camera~\cite{DAmen2021} has shown one can collect photons produced by alpha particles in a thin layer of LYSO scintillator~\cite{lyso} (Lutetium Yttrium Orthosilicate with chemical formula: $\textrm{Lu}_{2}\textrm{SiO}_{5}\textrm{:Ce}$). The same principles can be used to detect the scintillation photons produced in LYSO by the 2.73 MeV Tritium and the 2.05 MeV alpha particles. The decay angle of the two products is 180$^{o}$, thus only one (per thermal neutron) is ever scintillated into photons.
	
	The Timepix3 is a hybrid active pixel detector that consists of a semiconductor sensor bonded with the Timepix3~\cite{poikela2014timepix3} read-out ASIC. It has been well-researched and found uses in a wide range of applications in particle physics and life sciences. The key feature of the detector is the simultaneous measurement of the Time-Over-Threshold (ToT) and Time-of-Arrival (ToA) of a signal in each pixel, ToT is the energy lost by a particle in a pixel via ionisation and ToA is the time-stamp of when the hit was first registered. The chip has an almost dead-time free measurement up to hit rates of 80 MHits/s when operated in event measuring mode (self-triggered). The Timepix3 has a 256 $\times$ 256 (65,536 total) pixel matrix with each square pixel pitch measuring 55 \textmu m, covering a total area of 1.4 $\times$ 1.4 cm$^{2}$. Each hit has a nominal time resolution of 1.56 ns. TPX3Cam is an off-the-shelf version of the Timepix3 system based on SPIDR and developed by Amsterdam Scientific Instruments (ASI)~\cite{tpx3cam, Sen2020, Ianzano2020}, the silicon sensor coupled with it can image light of $> 10^{3}$ photons with $> 90\%$ quantum efficiency in the 400 - 1000 nm wavelength range \cite{Nomerotski2017}. To detect fewer photons, an image intensifier is required to multiply each photon into groups containing an over-threshold number of photons. The Cricket image intensifier is a micro-channel plate device developed by Photonis with an amplification factor of $>> 10^{3}$, and outputs 420 nm photons \cite{Photonis}. Single neutrons can be reconstructed by using the centre-of-mass (CoM) of ToT of the pixels activated by the cluster of photons \cite{timepixcam}, CoM is the peak of the reconstructed 2D Gaussian distribution of an event using the energy of each hit.
	
	The advantage of this approach to detect thermal neutrons optically is the converter and scintillator combination can be manufactured to cover a large area, and the resulting photons can be focused via lenses onto a small silicon sensor on a single TPX3Cam, which allows cost savings. The scintillation photons can be transmitted with radiation hard optical fibres or an array of mirrors in a light-tight enclosure, thus decoupling the delicate camera system from the scintillator to avoid direct neutron radiation. 
	
	\section{Experimental Setup}
	\label{sec:setup}
\begin{figure}[htbp]
		\centering
		\begin{subfigure}{\linewidth}
			\centering
			\includegraphics[width=\linewidth]{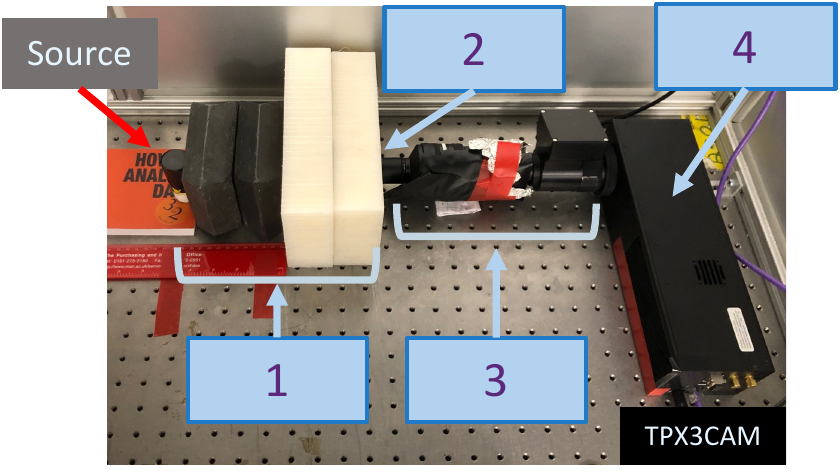}
			\caption{Photo of the setup.}
			\label{fig:setup1}
		\end{subfigure}
		\begin{subfigure}{\linewidth}
			\centering
			\includegraphics[width=\linewidth]{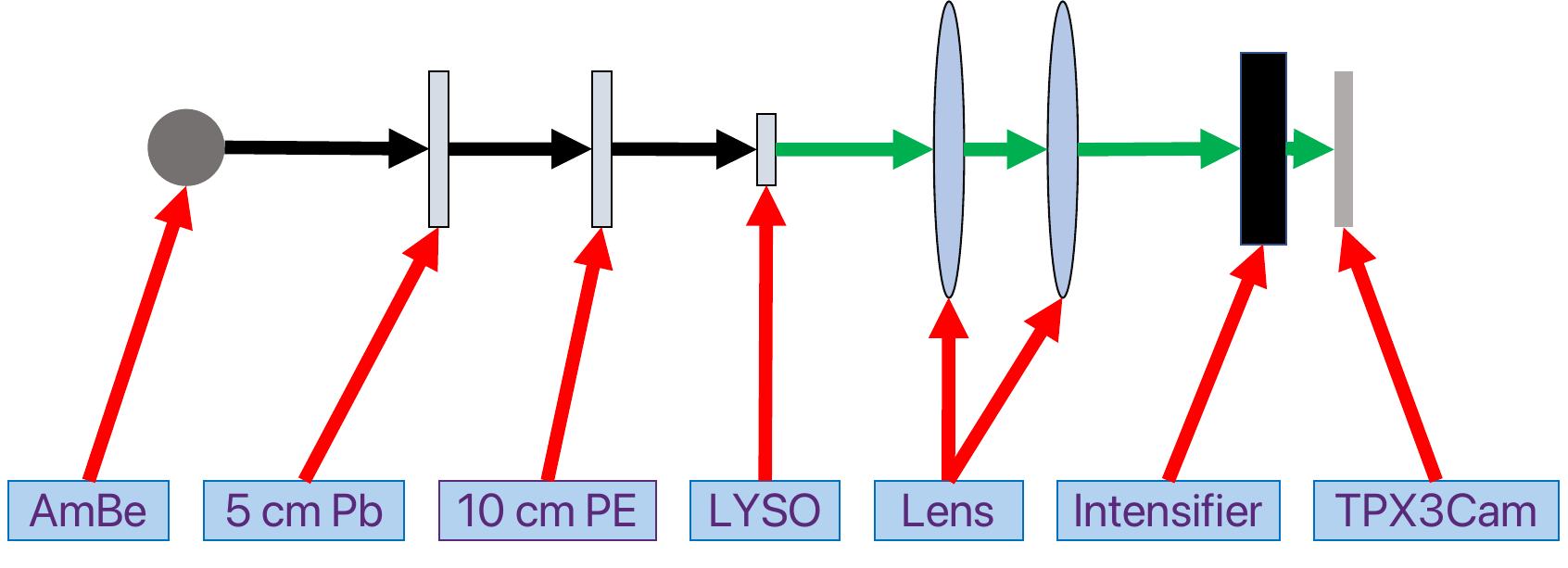}
			\caption{Illustration of the setup.}
			\label{fig:setup2}
		\end{subfigure}
		\caption{\ref{fig:setup1} shows the thermal neutron detection platform. 1 is the fast to thermal neutron moderation, 10 cm high-density polythene (PE), and gamma shielding, 5 or 10 cm lead (Pb). 2 is lithium-neutron reaction and photon scintillation, it is $^{6}$Li$_{2}$CO$_{3}$ of uneven thickness (10 to 100 \textmu m) deposited onto the surface of the 0.5 mm LYSO crystal, shown in figure~\ref{fig:lyso}. 3 is the pair of lenses of equivalent aperture f/2 which gave 1:1 magnification. The square component is the Cricket image intensifier. 4 is the photon detector TPX3Cam. In~\ref{fig:setup2}, the illustration of the setup is shown.}
		\label{fig:setup}
	\end{figure}

\begin{figure}[htbp]
		\centering 
		\includegraphics[width=.5\linewidth]{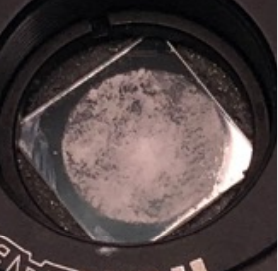}
		\caption{The LYSO crystal securely seated in the custom-made aluminium bracket, the white paste is made of $^{6}$Li$_{2}$CO$_{3}$, it was deposited via evaporation producing an uneven thickness of 10 to 100 \textmu m.}
		\label{fig:lyso}
	\end{figure}

	The light-tight thermal neutron detection platform is illustrated in figure~\ref{fig:setup}. Neutrons enter through the aluminium cap, reacting with approximately 10 - 100 um thick $^{6}$Li$_{2}$CO$_{3}$ paste distributed unevenly on the surface of the LYSO scintillator, measuring at 10 $\times$ 10 $\times$ 0.5 mm$^{3}$, as shown in figure~\ref{fig:lyso}. The 1 cm$^{2}$ area of the LYSO also dictates the field of view of the system. Photons are guided by a pair of lenses of equivalent aperture f/2, it has 1:1 magnification, focused by hand by the TPX3Cam output, this optical arrangement was the brightest in the inventory. The surviving scintillation photons enter the image intensifier to be multiplied. At the final detection stage, the green photons are detected by the TPX3Cam. The goal of the setup development was to have a compact system which can be used for remote monitoring away from the beam, that has spatial resolution, and its scintillator can be easily switched.
	
	The neutron source used in this experiment is a relatively old Americium-Beryllium (AmBe) capsule of activity 1.1$\times$10$^{6}$ n/s. In the setup with 5 cm of lead, the GEANT4 simulation shows $\sim$ 1.32 thermal neutron per second is converted and scintillated into photons by $^{6}$Li:LYSO. Since the actual source used is very old, its activity is reduced.
	
	\section{Simulation}
	\label{sec:sim}
	Simulations have been performed using Geant4 version 10.06 to identify the optimal converter thickness. The LYSO geometry has been set up in the simulation as 1 $\times$ 1 cm$^2$ area with a thickness of 500 \textmu m and an adjustable thickness of $^{6}$Li$_{2}$CO$_{3}$ covering the scintillation front side as a sensitive film for thermal neutron detection. 
	The simulation modelling assumes that neutron capture is distributed uniformly within the converter film. Each event begins with the generation of a random position within the sensitive film volume where a thermal neutron will be captured. From this point, an isotropic source with an arbitrary direction has been assigned where both alpha particles with an energy of 2.05 MeV and tritium particles with an energy of 2.73 MeV are generated in the opposite direction, transfer the converter layer and deposited the energy in the sensor volume. Then the thermal neutron detection efficiency $\varepsilon_n$ of the detector can be calculated using equation~\ref{Neutron Efficiency Equation}.
	
	\begin{equation}
		\label{Neutron Efficiency Equation}
		\varepsilon_n = \frac{n}{N} \times P(x) = \frac{n}{N} \times [1 - exp(\frac{-N_A}{\omega_A} \times \rho \sigma x)] 
	\end{equation}
	
\begin{figure}
		\centering
		\includegraphics[width = \linewidth]{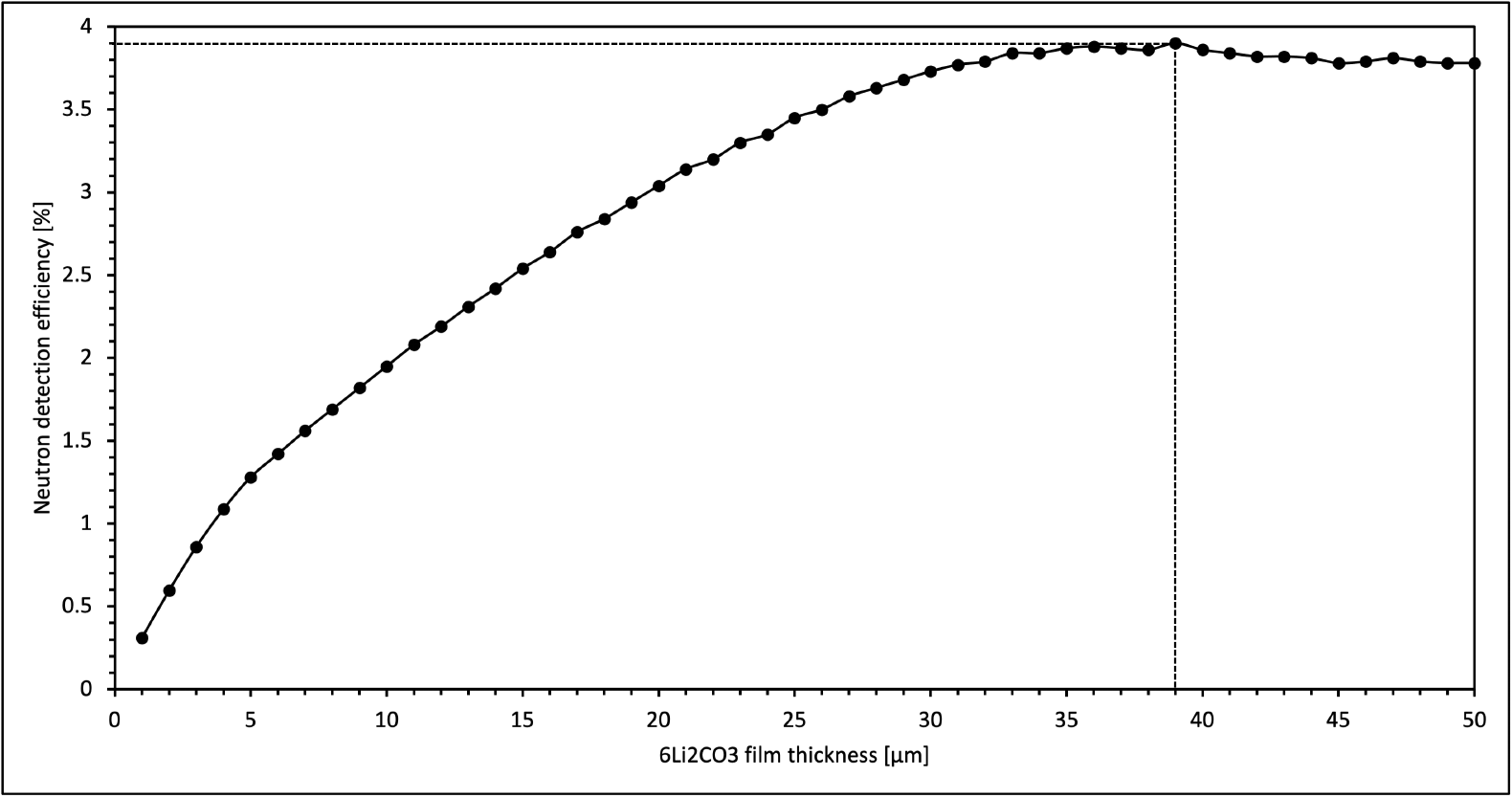}
		\caption {The dependency of the neutron detection efficiency as a function of the thickness of the neutron converter material, where the dashed line shows the maximum detection efficiency at 4\% with 39$\mu$m converter thick.}
		\label{fig:Geant4_model}
	\end{figure}

	Where $n$ is the number of detected events, $N$ the total number of generated events within the converter, $P(x)$ is the probability of thermal neutron absorption depending on the converter thickness $x$, $N_A$ is Avogadro’s number (6.022 $\times$ 10$^{23}$/mol), $\omega_A$ the atomic or molecular weight of the reactive film (for $^{6}$Li$_{2}$CO$_{3}$ is 72g/mol), $\rho$ the density of the reactive film (for $^{6}$Li$_{2}$CO$_{3}$ is 2.11g/cm$^3$) and $\sigma$ the thermal neutron cross-section (for $^6$Li is 940 b).
	The modelling of thermal neutron detection efficiency for a range of $^{6}$Li$_{2}$CO$_{3}$ film responsive thicknesses is present in figure~\ref{fig:Geant4_model}, which shows variation in detection efficiencies depending on selection of the neutron converter thickness. The neutron detection efficiency can be calculated when a tritium or an alpha particle can reach the sensor that is identified as a neutron conversion coincident event. Where the maximum detection efficiency of a single event is around 4\% with 39$\mu$m thick. Increasing the converter thickness will increase the neutron capture probability, as well as the number of tritium or alpha particles absorbed. Thus the detection efficiency curve saturates after the 39$\mu$m.
	
	\section{Results and Analysis} 
	\label{sec:analysis}
	
	\subsection{Detection efficiency}
	\label{sec:Detectionefficienc}
	
	The conversion efficiency of $^{6}$Li is $\sim$ 4\% at 39$\mu$m. Every tritium/alpha can be seen by the TPX3Cam. The steps from scintillation to detection of the system are as follows:
	\begin{enumerate}
		\item The LYSO scintillator has a photon yield of 27,600 photons/MeV~\cite{lyso}, for 2.05 MeV and 2.73 MeV it produces approximately 56,600 and 75,300 photons respectively.
		\item When the particle releases energy into the crystal, the quenching effect suppresses the photon yield by a factor of 11~\cite{KOBA2011, AHMADOV2013, vonKrosigk2016, Ogawa2018, YANAGIDA2018, DAmen2021}.
		\item refraction occurs between the crystal and air interface (refraction indexes of 1.82 and 1.00 respectively), the distance between LYSO and the intensifier input is 100 mm, and the solid angle reduces the number of photons by a factor of 18.
		\item The efficiency is 0.5\% at the input of the intensifier, which itself has a photon detection efficiency of 20\%.
	\end{enumerate}
	Taking all these factors into account, for each successfully converted thermal neutron, the TPX3Cam sees $\sim$ 5-7 photon groups (each photon is multiplied into a group by the intensifier) arriving at the camera. This can be further reduced by effects such as reflection on the internal wall of the lenses, optical attenuation, and hit blending.
	
	\subsection{Event Reconstruction and identification}
	\label{sec:EventReconstructionandidentification}
	
\begin{figure}
		\centering
		\includegraphics[width = 0.8\linewidth]{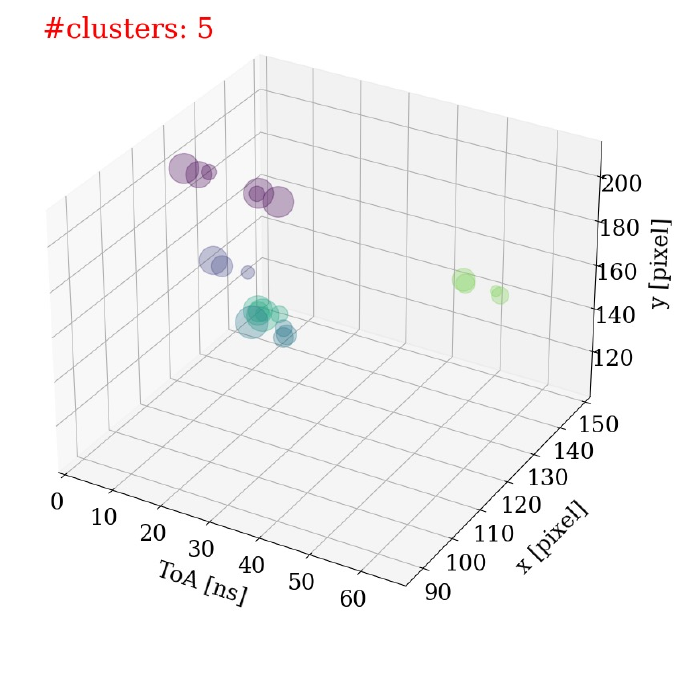}
		\caption {Signals reconstructed from TPX3Cam during 70 ns slice of time. The x and y axis represents the coordinate of the pixels of the detector. ToA is the time stamp of a pixel hit. Each circle is the over-threshold signal produced by a group of photons outputted by the image intensifier, of which the size is the ToT of the hit and the colour shows that they are likely from the same parent photon before multiplication.}
		\label{fig:eventexample}
	\end{figure}

	The identification of a neutron signal (or rather the product of the $^{6}$Li(n,$\alpha$) reaction) can be done by clustering a specific number of photons that would arrive at the camera within the decay time of the LYSO crystal, as shown in figure~\ref{fig:eventexample}.
	
\begin{figure}[htbp]
		\centering
		\begin{subfigure}{\linewidth}
			\centering
			\includegraphics[width=\linewidth]{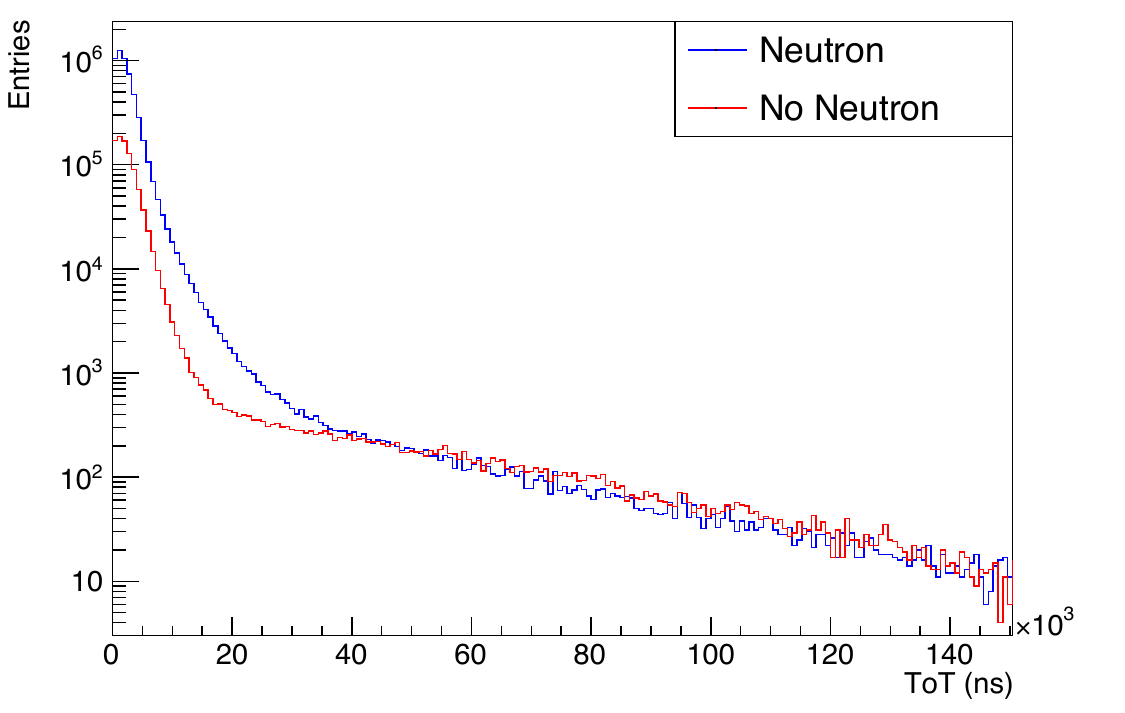}
			\caption{Integrated ToT}
			\label{fig:analysisplots1}
		\end{subfigure}
		\begin{subfigure}{\linewidth}
			\centering
			\includegraphics[width=\linewidth]{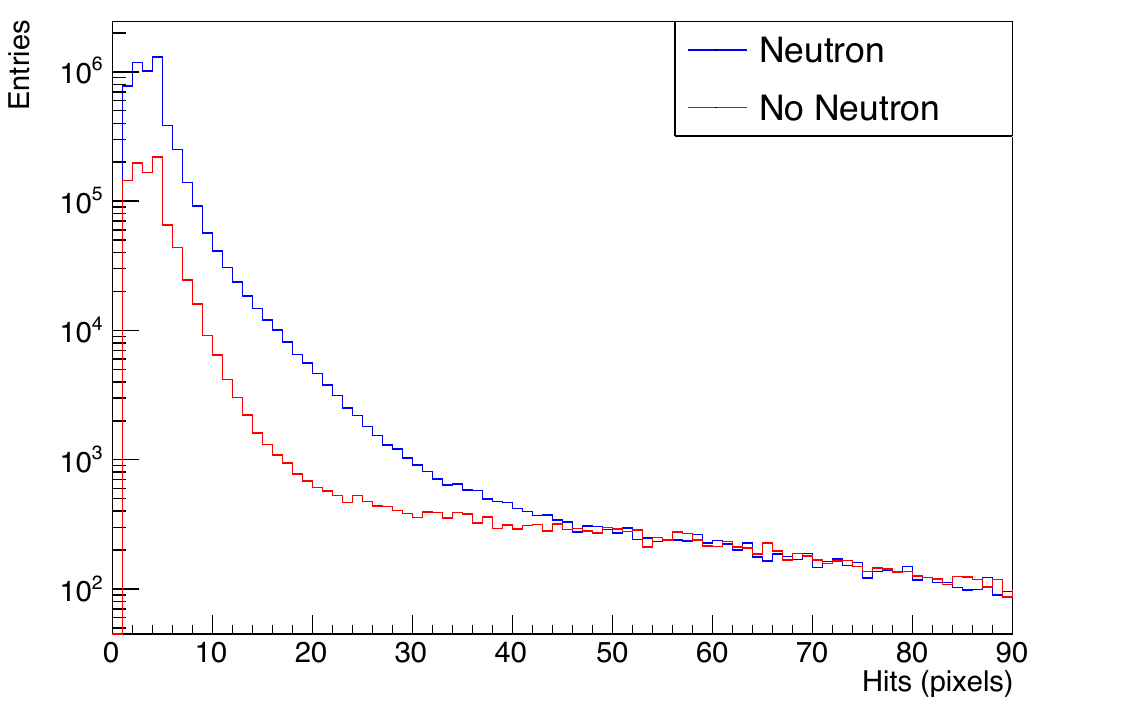}
			\caption{Hits per event}
			\label{fig:analysisplots2}
		\end{subfigure}
		\begin{subfigure}{\linewidth}
			\centering
			\hspace*{-0.60cm}\includegraphics[width=0.93\linewidth]{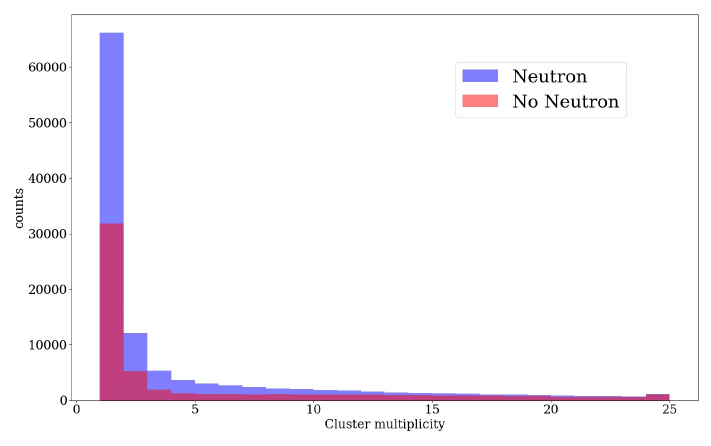}
			\caption{Clusters per event}
			\label{fig:analysisplots3}
		\end{subfigure}
		\caption{Plots showing the integrated ToT, this is equivalent to the total energy of an event. It is correlated to the total number of photons (top). The number of hits is the total number of ``hits'' in an event (middle). The number of clusters is the total number of scintillation photons seen by the TPX3Cam from LYSO of an event (bottom). The legend ``Neutron'' and ``No Neutron'' means with and without the AmBe source respectively.}
		\label{fig:analysisplots}
	\end{figure}

	Each ``hit'' is a group of photons ($\gamma_{multi}$) emitted by the intensifier, represented by a single dot in figure~\ref{fig:eventexample}. Multiple hits within a specific spatial and temporal window are grouped to form a ``cluster'', which is the reconstruction of a single scintillation photon ($\gamma_{scint}$) before being multiplied by the intensifier, these hits have the same colour in figure~\ref{fig:eventexample}. Multiple ``clusters'' are grouped to form an ``event'', which is the reconstruction of the calorimetric signal of a charged particle depositing its energy in LYSO. This gives three simple variables:
	\begin{enumerate}
		\item Integrated ToT (illustrated in figure~\ref{fig:analysisplots1}), this is equivalent to the total energy of an event. It is correlated to the total number of photons ($\gamma_{multi}$).
		\item Hits per event (illustrated in figure~\ref{fig:analysisplots2}), this is the total number of hits in an event. Photons ($\gamma_{multi}$) could trigger multiple pixel hits, thus it is not always proportional to the integrated ToT.
		\item Clusters per event (illustrated in figure~\ref{fig:analysisplots3}), this is the total number of scintillation photons seen by the TPX3Cam from LYSO of an event. This discriminates particles deposited more/less energy than the signal tritium/alpha into the LYSO crystal.
	\end{enumerate}
	These parameters were chosen in a bid to discriminate the signal from the backgrounds. Other parameters such as the delta ToA between hits and clusters, and others were ineffective in their discrimination ability.
	
	The backgrounds include MeV gamma-rays, the simulation shows $> 90\%$ is attenuated by 5 cm of lead block. The LYSO crystal is a beta emitter, $^{176}$Lu. When powered on, the image intensifier produces dark counts with no inputs. Together these factors will make up a portion of the background in this experiment. The rate of which is shown in TAB~\ref{tab:ratetable}.
	
	The chosen filtering parameters are:
	\begin{enumerate}
		\item Integrated ToT, 8x10$^{3}$ $<$ x $<$ 40x10$^{3}$ (ns).
		\item Number of hits per event, 15 $<$ x $<$ 50.
		\item Number of clusters per event, 3 $<$ x $<$ 8. As calculated in the previous section, each thermal neutron can convert to a maximum of 5-7 photon groups.
	\end{enumerate}
	
	Further analysis can be done to find the CoM of an event, it can approximate the coordinate of the neutron product n in the LYSO crystal, thus making imaging possible. This will be investigated as the next step of the project.
	
	\section{Discussion}
	\label{sec:discussion}
	Simulation gives a neutron rate of $\sim$ 1.32 Hz at the TPX3Cam. The experimental results of the event frequency with a range of source activity (by changing the distance between the source and the scintillator) before and after the event filtering was applied are shown in TAB~\ref{tab:ratetable}. In the after-dataset, by taking away the backgrounds, the ratio between the datasets approximates their expected value. The thermal neutron rate also agrees with the simulation. 
	
	\begin{table}[ht!]
		\centering
		\caption{The event frequency/rate before and after event filtering were applied to each experimental dataset, all rates are in Hz. The ratio column is the ratio of the current rate to the maximum rate.}
		\label{tab:ratetable}
		\begin{subtable}{\linewidth}
			\centering
			\begin{tabular}{|l|l|l|l|}
				\hline
				Source&Reconstructed events&Minus Background&Ratio\\ \hline
				Background&34.2$\pm$0.38&0$\pm$0.38&-\\ \hline
				Max-rate&67.5$\pm$0.42&33.3$\pm$0.42&1\\ \hline
				Half-rate&52.5$\pm$0.41&18.3$\pm$0.41&0.55\\ \hline
				Quarter-rate&40.8$\pm$0.42&6.6$\pm$0.42&0.2\\ \hline
			\end{tabular}
			\caption{Before event filtering applied.}
			\label{fig:beforetable}
		\end{subtable}
		\begin{subtable}{\linewidth}
			\centering
			\begin{tabular}{|l|l|l|l|}
				\hline
				Source&Reconstructed events&Minus Background&Ratio\\ \hline
				Background&0.8$\pm$0.01&0&-\\ \hline
				Max-rate&2.0$\pm$0.01&1.2$\pm$0.01&1\\ \hline
				Half-rate&1.5$\pm$0.01&0.7$\pm$0.01&0.58\\ \hline
				Quarter-rate&1.1$\pm$0.01&0.3$\pm$0.01&0.25\\ \hline
			\end{tabular}
			\caption{After event filtering applied.}
			\label{fig:aftertable}
		\end{subtable}
	\end{table}
	
	The camera system can detect thermal neutrons by subtracting the background from the total flux from the source, but the current setup cannot pinpoint thermal neutron events with a high confidence level due to two factors. First, the LYSO crystal does not have a gamma-neutron discrimination mechanism. Second, an array of charged particles produced inside of the LYSO or the $^{6}$Li adds to the background, as the system was fully enclosed in aluminium, shielding external radiation from polluting the measurement.
	
\begin{figure}
		\centering
		\includegraphics[width = \linewidth]{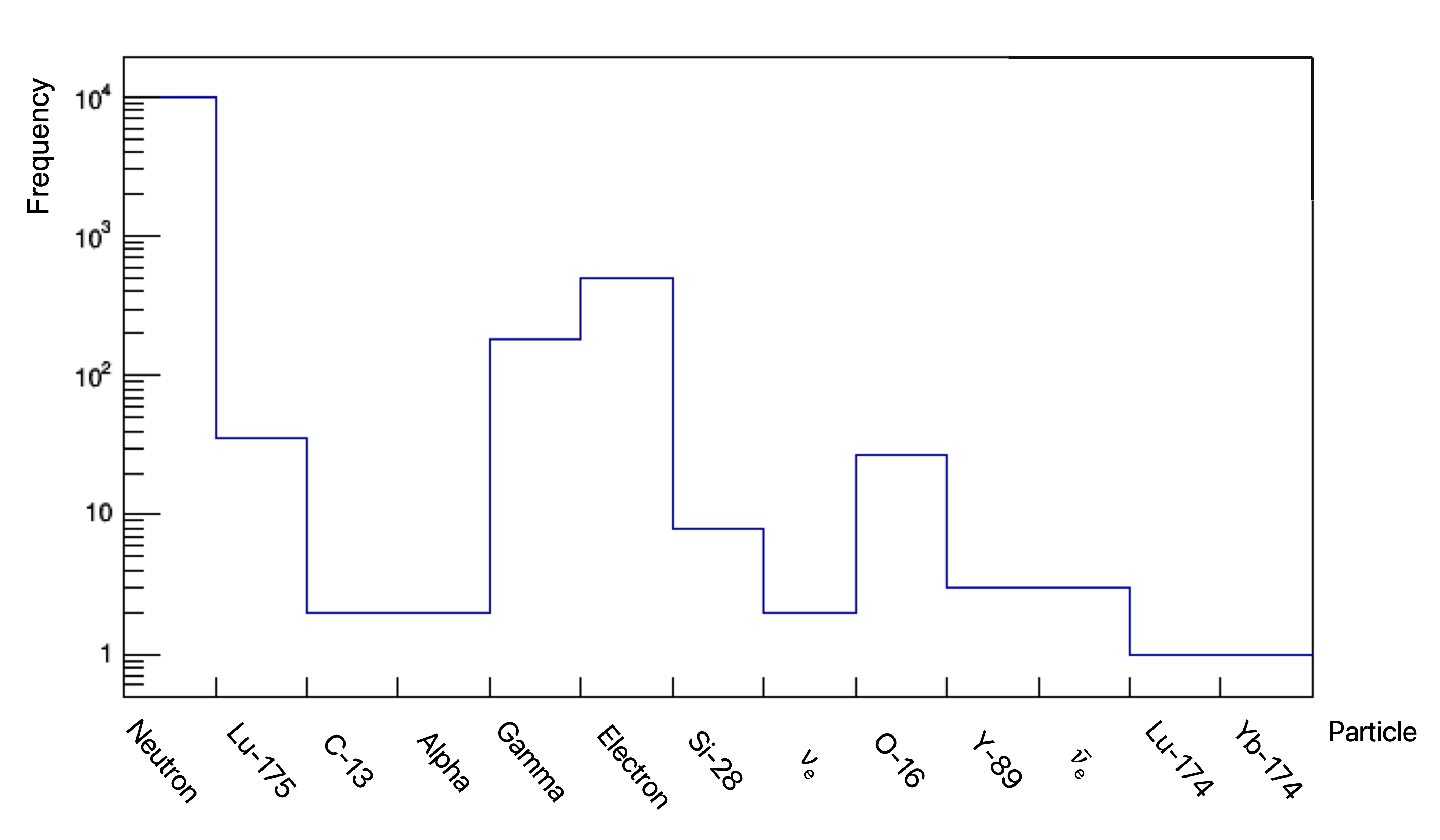}
		\caption {Simulation results of all particles detected inside of the $^{6}$Li:LYSO scintillator, note: neutron is the flux from the AmBe source and the neutron conversion to tritium and alpha was not simulated. The charged particles would not have been able to penetrate the 2 mm aluminium enclosure, they have to be produced inside of the scintillator.}
		\label{fig:particles}
	\end{figure}

	To further investigate the second factor, the overall detected flux of all particles was calculated and compared against the simulated data in TAB~\ref{tab:efficiencytables}. Americium-Beryllium source has a 50-50 fast neutron ($\sim$ 3 MeV) and gamma (0.2 to 4.4 MeV) production. The lead and polythene blocks were simulated to reduce the overall neutron and gamma flux to 9\%. However, the experimental results show that 78\% of the overall flux is still seen by the detector. We theorize this is due to charged particles produced inside of the LYSO crystal, this was confirmed by the simulation shown in figure~\ref{fig:particles}. TAB~\ref{tab:efficiencytables} indicate we are sensitive to thermal neutrons by the increased number of event rates (compared to only $^6$Li) after the PE block was added.
	
	\begin{table}[htbp]
		\centering
		\caption{The total flux of both signal and background particles arriving at the surface of the LYSO crystal due to the addition of the lead and polythene blocks. Note the increase of events with polythene blocks versus no blocks, and PE + PB versus Pb only. These extra events are the thermal neutrons.}
		\label{tab:efficiencytables}
		\begin{tabular}{|l|l|l|l|l|l|}
			\hline
			\begin{tabular}{@{}c@{}}Observed\\Events\end{tabular}&\begin{tabular}{@{}c@{}}No block\\only $^{6}$Li\end{tabular}&\begin{tabular}{@{}c@{}}Lead\\block\end{tabular}&\begin{tabular}{@{}c@{}}Polythene\\block\end{tabular}&\begin{tabular}{@{}c@{}}PE+PB\\blocks\end{tabular}&\begin{tabular}{@{}c@{}}No\\AmBe\end{tabular}\\ \hline
			Experimental&100\%&77.94\%&106.86\%&78.13\%&9\%\\ \hline
			Simulation&100\%&50.3\%&46\%&9\%&0\%\\ \hline
		\end{tabular}
	\end{table}
	
	\section{Conclusion}
	\label{sec:conclusion}
	We used a Timepix3-based optical event camera to measure thermal neutrons in real-time with a spatial resolution of $\sim$ 16 \textmu m and a temporal resolution of 1.56 ns. 
	
	The source used is Americium-Beryllium with 1.1$\times$10$^{6}$ n/s omnidirectional flux, equivalent to 389 n/s reaching the converter surface. In the simulation, the system of 5 cm of lead, 10cm high-density polythene, and $\sim$ 10 um $^{6}$Li converter has a combined efficiency of $\sim$ 0.34\%, it predicts a detectable thermal neutrons rate of 1.32 Hz. 
	
	The experiment results show 5 cm of lead and 10 cm of polythene each only suppressed 20\% and 0\% total flux respectively, did not achieve the 90\% flux reduction as simulated. However, the chosen filtering parameters were able to effectively suppress the additional background, giving a signal rate of 1.2 Hz, which is in line with the simulation. 
	
	We conclude that thermal neutrons were successfully detected using a $^{6}$Li:LYSO neutron converter scintillator combination and the TPX3Cam as the detector. The chosen filtering parameters were effective in suppressing the main background. However, neutron imaging is not possible with LYSO due to its lack of neutron and gamma discrimination mechanism.
	
	If neutron moderation is not needed, i.e. in a primary thermal neutron field, the system efficiency increases to 4\%, which is the efficiency of the converter. Similarly, if the incoming neutron energy is higher, a thicker moderation layer is required which will decrease the overall efficiency.
	
	Compared to directly detecting the ions from the interaction of thermal neutron and converter material, this optical approach gives the possibility of free-space light collection, it enables measurements to be conducted remotely, away from the radiation source. Radiation-hard optical fibres or enclosed mirrors can be used in long-distance applications to minimise photon collection losses. In the study done by~\cite{losko2021new}, the popular $^{6}$LiF:ZnS material was used to convert thermal neutrons to photons. In contrast, we used the novel non-standard neutron detector of $^{6}$LiF deposited onto the surface of LYSO, this combination is a fraction of the cost of the $^{6}$LiF:ZnS compound and this report shows that it can detect the presence of neutron in the background. The setup exhibits a low signal-to-noise ratio of 1/67.5 due to a large background, which consists of gamma radiation from the AmBe source and charged particles produced inside the scintillator via neutron elastic collision.
	
	This system with the custom bracket and light-tight photon multiplication channel (figure~\ref{fig:setup} and \ref{fig:lyso}) is set up to be a test platform for a range of different neutron scintillators, such as the compounds EJ-420 and EJ-204 reported in~\cite{manfredi2020single}, to find the most cost-effective and bright candidate suitable for the next stage of the project - a portable thermal neutron imager. All detection sub-components of the system have position sensitivity (scintillator, intensifier, and TPX3Cam). The current field-of-view is 1cm$^{2}$ as dictated by the size of the LYSO scintillator, it can be increased with minimal modification by accommodating a larger-sized scintillator coupled with a focusing lens, such as proposed in~\cite{Mavrokoridis2015} for detecting cosmic muons.
	
	\section{Acknowledgment}
	Special thanks to Dr Paul Campbell of the University of Manchester for providing much-needed nuclear materials and knowledge for the experiment.
	
	\bibliographystyle{IEEEtran}
	\bibliography{neutrons}
	
\end{document}